\documentclass[aps,prl,reprint,superscriptaddress,nofootinbib,longbibliography,floatfix]{revtex4-2}

\usepackage{graphicx}
\usepackage{amsmath,amssymb,amsfonts}
\usepackage{xcolor}
\usepackage{hyperref}

\graphicspath{{figures/}}
\newcommand{\nsat}{n_{\rm sat}}
\newcommand{\cstwo}{c_s^2}
\newcommand{\tLambda}{\tilde{\Lambda}}
\newcommand{\Msun}{M_\odot}
\newcommand{\nrise}{n_{\rm rise}}
\newcommand{\cslow}{c_{s,{\rm low}}^2}
\newcommand{\cspeak}{c_{s,{\rm peak}}^2}
\newcommand{\cshigh}{c_{s,{\rm high}}^2}

\begin{document}

\title{Which Neutron Stars Reach the Stiffening Regime? Multimessenger Constraints on Core Sound Speed and Stellar-Mass Thresholds}

\author{Nicol\'as Viaux}
\affiliation{Departamento de F\'isica, Universidad T\'ecnica Federico Santa Mar\'ia, Casilla 110-V, Valpara\'iso, Chile}
\affiliation{Millennium Institute for Subatomic Physics at the High-Energy Frontier (SAPHIR), Santiago, Chile}

\author{Sebasti\'an Mendizabal}
\affiliation{Departamento de F\'isica, Universidad T\'ecnica Federico Santa Mar\'ia, Casilla 110-V, Valpara\'iso, Chile}

\begin{abstract}
We present a concise multimessenger inference of the neutron-star core
sound-speed profile using GW170817 and three \textit{NICER}
mass--radius posteriors (PSR J0030$+$0451, PSR J0740$+$6620, and
PSR J0437$-$4715). The main result is not only a preference for
intermediate-density stiffening within smooth equation-of-state families,
but a translation of that inference into the stellar masses that access
the relevant density regime. In the baseline smooth peaked family, the
posterior probability that $\cstwo > 1/3$ at $3.5\,\nsat$ is $85.4\,\%$,
while equal-prior averaging over peaked, monotonic, piecewise, and
transition-capable families gives a more conservative
$79.0\,\%$. Posterior-resampled exact Tolman--Oppenheimer--Volkoff
solutions show that the onset density of the inferred stiffening is
typically reached near $1.6\,\Msun$, whereas the peak region is accessed
only near $2.1\,\Msun$. A J0740-like $2.07\,\Msun$ pulsar reaches the
onset in $91\,\%$ of posterior draws but the peak in only $46\,\%$,
showing that current data mainly constrain whether massive stars have
entered the stiffening regime rather than traversed its full peak.
\end{abstract}

\maketitle

\noindent\textit{Electronic address:} \href{mailto:nicolas.viaux@usm.cl}{nicolas.viaux@usm.cl}

\noindent\textit{Electronic address:} \href{mailto:sebastian.mendizabal@usm.cl}{sebastian.mendizabal@usm.cl}

The speed of sound in dense matter,
$\cstwo \equiv \partial P/\partial \epsilon$, provides a direct probe of
how strongly the neutron-star equation of state (EOS) stiffens above
nuclear saturation density. At asymptotically high density, perturbative
QCD implies $\cstwo \to 1/3$ from below~\cite{Kurkela2010,Fraga2014,Gorda2023},
whereas several multimessenger analyses have argued that neutron-star
cores may exceed that conformal value at intermediate densities
inside observed stars~\cite{Annala2018,Annala2023,Brandes2023}. The key
question is therefore not only whether supra-conformal sound speeds are
favored, but which observed neutron stars actually probe the relevant
core-density regime.

We address that question by combining the tidal information from
GW170817~\cite{Abbott2017,Abbott2018tidal,Abbott2019eos} with
\textit{NICER} mass--radius posteriors for
PSR J0030$+$0451~\cite{Riley2019,Miller2019,Vinciguerra2024},
PSR J0740$+$6620~\cite{Riley2021,Miller2021}, and
PSR J0437$-$4715~\cite{Choudhury2024}, together with the requirement that
the EOS support radio pulsars above $2\,\Msun$~\cite{Demorest2010,Antoniadis2013,Cromartie2020,Fonseca2021}.
These data probe the density range reached in neutron-star cores and
therefore the range in which any strong stiffening or transition-like
softening must occur if it is to influence observable masses, radii, and
tidal deformabilities.

Our baseline EOS family is a smooth five-parameter sound-speed profile
constructed to make the onset density $\nrise$, the transition width
$\delta_n$, and the peak height $\cspeak$ direct inference targets. From each sampled
$c_s^2(n)$ profile we construct the EOS, solve the
Tolman--Oppenheimer--Volkoff and tidal equations, and evaluate the
multimessenger likelihood. The production run is accelerated by a
machine-learning emulator trained on $7{,}256$ exact stellar models, and
the main posterior results are validated with exact posterior-resampled
TOV solutions. Details of the broader framework, robustness tests, and
cross-family comparison are given in the companion long-form analysis.
The profile is written as
\begin{equation}
\begin{aligned}
\cstwo(n)={}& \cslow
+ \frac{\cspeak-\cslow}{2}
\left[1+\tanh\!\left(\frac{n-\nrise}{\delta_n}\right)\right] \\
&+ \frac{\cshigh-\cspeak}{2}
\left[1+\tanh\!\left(\frac{n-\nrise}{\delta_n}-2\right)\right],
\end{aligned}
\label{eq:prl_profile}
\end{equation}
which makes explicit the density at which the sound speed starts to rise,
the width of that rise, the peak value reached near
$\nrise+\delta_n$, and the high-density relaxation toward
$\cshigh$. In the present analysis this analytic profile is used only as
an inference scaffold: it is constrained by the data, translated into a
thermodynamically consistent EOS, and then checked against less
restrictive monotonic, piecewise, and transition-capable families.

The multimessenger likelihood is built directly from public posterior
samples rather than Gaussian contour approximations~\cite{Abbott2019eos,Riley2021,Choudhury2024,Vinciguerra2024}. For GW170817 we use
a two-dimensional kernel-density estimate in the
$(\mathcal{M}_c,\tLambda)$ plane based on the public LIGO--Virgo
posterior~\cite{Abbott2019eos}. For each \textit{NICER} source we use a
sample-based kernel-density estimate in the $(M,R)$ plane and evaluate
the EOS likelihood by averaging the posterior density along the model
curve $R(M)$. This allows the inference to retain the non-Gaussian
structure of the public X-ray posteriors while remaining computationally
tractable over a large proposal set.
The production run uses $10^5$ prior proposals and achieves an
effective sample size of $5.8\times10^4$, which is high for EOS
importance sampling because the five-parameter profile already
concentrates support in the physically relevant rise--peak--fall sector.
We further checked that emulator errors do not bias the conclusions by
recomputing exact TOV solutions for the highest-weight posterior
samples; the induced changes in posterior weights and in the headline
supra-conformal probability are negligible at the precision relevant
here.

\begin{figure}[t]
  \centering
  \includegraphics[width=\columnwidth]{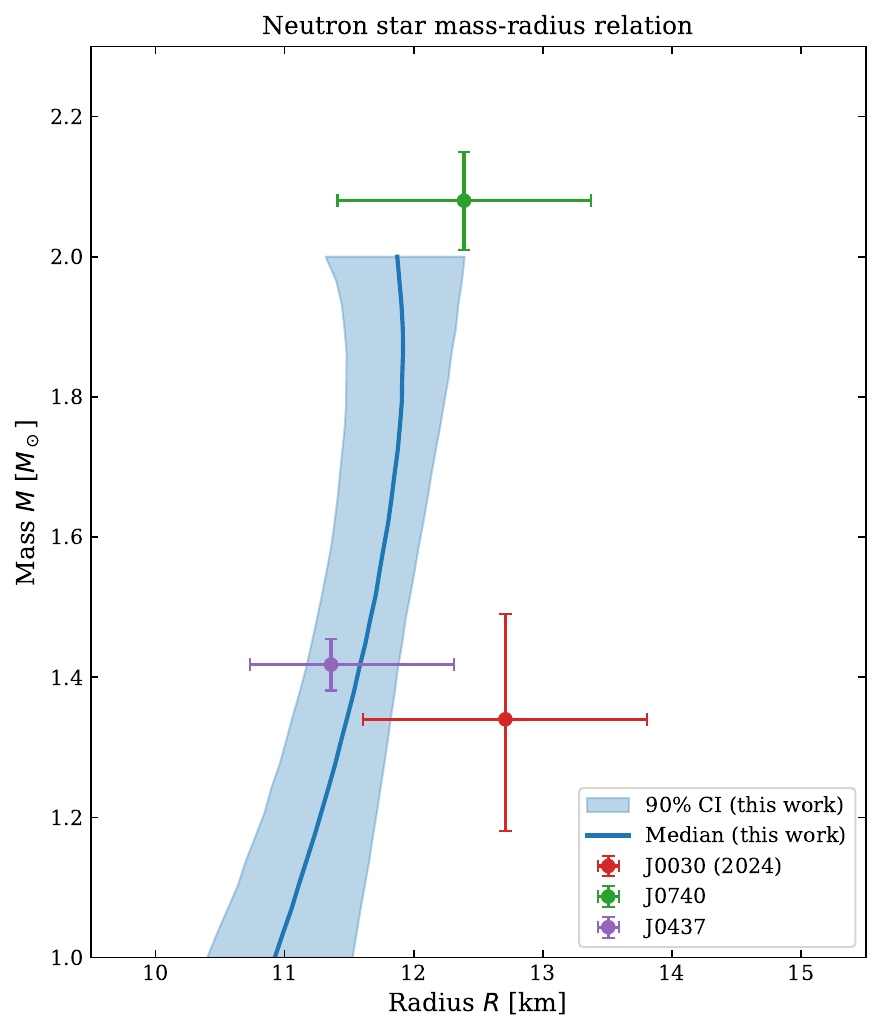}
  \caption{%
    Posterior 90\% credible band for the mass--radius relation,
    together with the public \textit{NICER} mass--radius constraints for
    PSR J0030$+$0451, PSR J0740$+$6620, and PSR J0437$-$4715.
    The band shows that the inferred EOS remains compatible with the
    observed radii while extending to the high-mass regime where the
    stiffening onset becomes observationally accessible.}
  \label{fig:mr}
\end{figure}

The observational anchor for the threshold interpretation is the
mass--radius band in Fig.~\ref{fig:mr}. The credible band passes through
the public \textit{NICER} constraints while extending smoothly from
canonical-mass stars to the high-mass J0740 regime. This matters
because the onset inference is not being driven by an isolated
high-density extrapolation: it is tied continuously to the same
posterior mass--radius band that fits the measured stars. In that
sense, the mass--radius relation shows where the dataset anchors the EOS
globally.

\begin{figure}[t]
  \centering
  \includegraphics[width=\columnwidth]{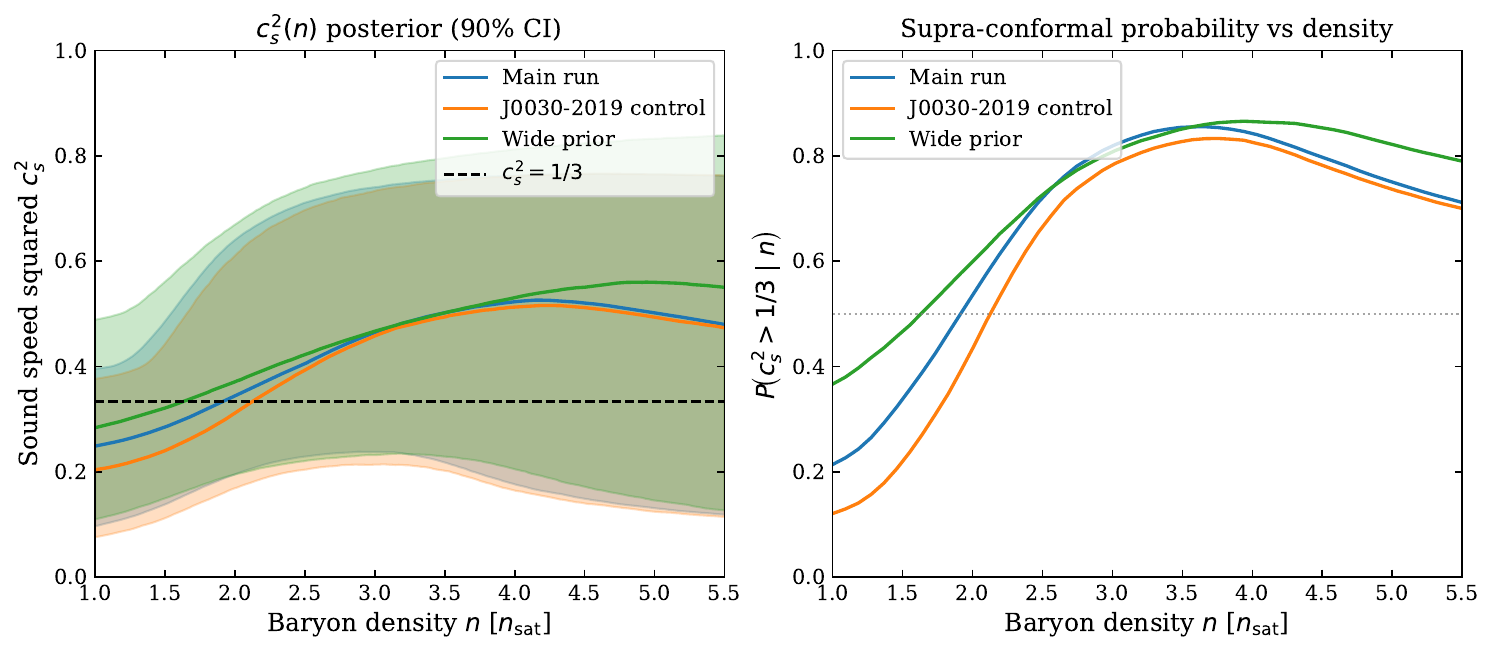}
  \caption{%
    Posterior median and 90\% credible band for $\cstwo(n)$ together
    with the posterior probability that $\cstwo>1/3$ as a function of
    density. The broad maximum around $3$--$4\,\nsat$ identifies the
    density interval in which smooth EOS families most strongly favor
    supra-conformal behavior.}
  \label{fig:cs2prob}
\end{figure}

The first headline result is that current data favor intermediate-density
stiffening within smooth EOS families, but not yet decisively. In the
baseline smooth peaked family, the posterior probability that
$\cstwo > 1/3$ at $3.5\,\nsat$ is $85.4\,\%$, as visualized in
Fig.~\ref{fig:cs2prob}. A monotonic smooth family
gives $95\,\%$, a six-knot piecewise family gives
$83^{+4}_{-4}\,\%$, and a transition-capable family that allows strong
intermediate-density softening gives $55\,\%$. Equal-prior averaging
over these four families yields a model-averaged supra-conformal
probability of $79.0\,\%$. The data therefore favor stiffening in the
few-$\nsat$ regime, but the precise strength of that statement remains
model-dependent once transition-like softening is admitted.
This point matters because the smooth peaked prior is not agnostic. In a
wide-prior variant that lowers the floor on the peak parameter down to
the conformal value, the prior probability for
$\cstwo > 1/3$ at $3.5\,\nsat$ is already $\approx 79\,\%$, and the
posterior rises only to $85.1\,\%$. The corresponding update is real but
modest: current data favor intermediate-density stiffening, yet they do
not by themselves establish it independently of EOS-family assumptions.
The most conservative compact summary is therefore the cross-family
model-averaged probability rather than the single-family headline value.

The second, and astrophysically more direct, result is the translation
from density space into stellar-mass space. In the baseline posterior,
the onset density is
$\nrise = 2.43^{+1.52}_{-1.00}\,\nsat$ and corresponds to a threshold
mass
$M_{\rm rise} = 1.59\,\Msun$ (90\% CI: $0.72$--$2.14\,\Msun$).
The peak-density proxy is reached at
$M_{\rm peak} = 2.08\,\Msun$ (90\% CI: $1.47$--$2.46\,\Msun$).
This means that the onset of strong stiffening is typically reached in
massive but common neutron stars, whereas the peak region is probed only
near the upper end of the currently observed mass distribution.
The posterior median chirp-mass-weighted tidal deformability is
$\tLambda = 205^{+38}_{-25}$, and the inferred radius at
$1.4\,\Msun$ in the baseline analysis is
$R_{1.4}=12.1^{+1.1}_{-1.3}\,{\rm km}$, consistent with recent
multimessenger EOS reconstructions. The astrophysical novelty is that
these global EOS constraints can now be converted into a statement about
which observed stars actually reach the core-density interval where the
stiffening occurs.

The density-space statement underlying the mass-threshold inference is
shown in Fig.~\ref{fig:cs2prob}. The posterior median of $\cstwo(n)$ rises
through the conformal value near a few times $\nsat$, and the
probability $P(\cstwo>1/3)$ develops a broad plateau across
approximately $3$--$4\,\nsat$ rather than a narrow spike at one special
density. This is why we quote $3.5\,\nsat$ as a representative headline
point: it lies within the broad maximum of the posterior support and is
close to the inferred peak-access regime of the heaviest observed stars.
The same figure also clarifies why the high-density asymptote remains
weakly constrained by present data: the observations probe the rise and
peak region more strongly than the eventual relaxation back toward the
conformal limit.
In other words, current multimessenger measurements determine where the
EOS stiffens much better than they determine how the sound speed
behaves far above the densities reached in presently observed stars.

\begin{figure}[t]
  \centering
  \includegraphics[width=\columnwidth]{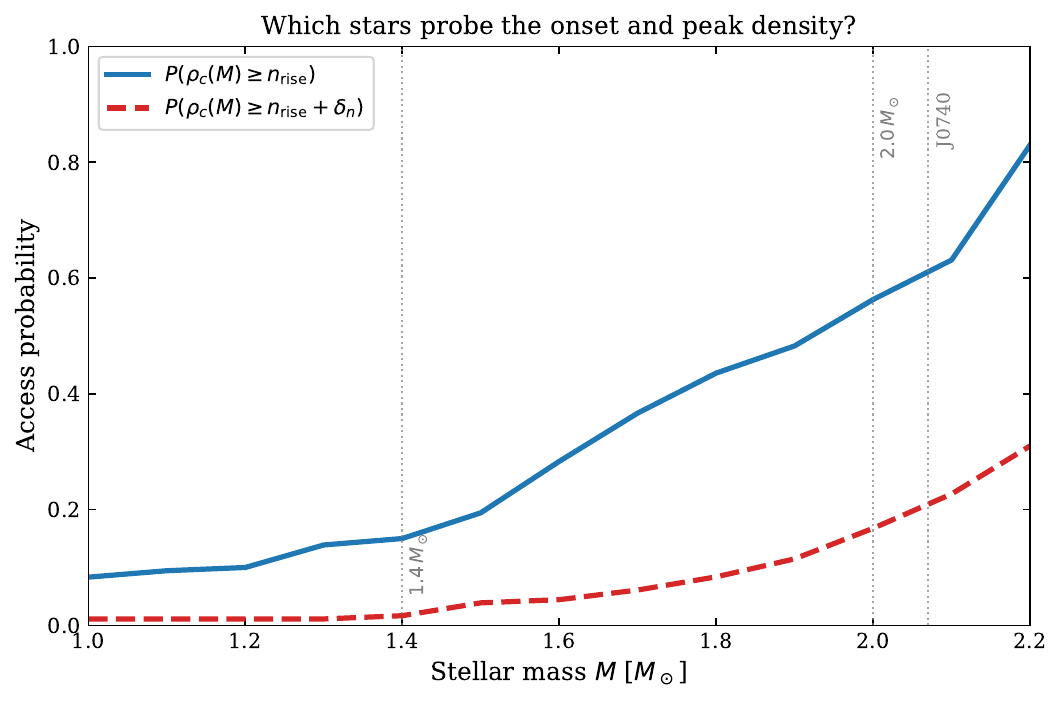}
  \caption{%
    Posterior-resampled exact-TOV access probabilities for the onset
    density $\nrise$ (solid) and the peak-density proxy
    $\nrise+\delta_n$ (dashed) as functions of stellar mass.
    A J0740-like pulsar is very likely to have entered the inferred
    stiffening regime, but is not yet guaranteed to probe its peak.
    This makes the figure the most direct observational summary of the
    present inference.}
  \label{fig:impact}
\end{figure}

The most important observational consequence appears in
Fig.~\ref{fig:impact}. The probability that a star reaches the onset density rises
rapidly with mass, while the probability of reaching the peak remains
below $50\,\%$ even for a J0740-like pulsar. Quantitatively, a
$2.07\,\Msun$ star reaches $\nrise$ in $91\,\%$ of posterior draws but
reaches the peak proxy $\nrise+\delta_n$ in only $46\,\%$. By contrast,
a canonical $1.4\,\Msun$ star reaches the onset in only $34\,\%$ of
draws and the peak in just $2\,\%$. Present multimessenger constraints
are therefore driven mainly by whether the most massive observed stars
have entered the stiffening regime, not by stars that fully traverse its
peak.
This provides a useful reinterpretation of the present observational
landscape. Canonical-mass stars are sensitive mainly to the low- and
intermediate-density part of the EOS and therefore only indirectly to
the stiffening onset, whereas the heaviest radio and X-ray pulsars
determine whether the stiffening region lies inside the currently
accessible stellar population. In that sense, J0740-like systems are the
critical bridge between abstract density-space inference and direct
astrophysical observables.
The comparison among EOS families sharpens that interpretation. The
monotonic family shows that a smooth EOS without a post-peak relaxation
can still give a strong preference for supra-conformal behavior at a few
times $\nsat$. The piecewise family shows that the result survives a
more agnostic smooth interpolation. The transition-capable family
identifies the main loophole: if the EOS is allowed to undergo a strong
intermediate-density softening dip before restiffening, then the
inference for $\cstwo>1/3$ weakens substantially even though the model
remains statistically competitive with the smooth families.

This interpretation sharpens the usual statement that current data favor
$\cstwo > 1/3$ somewhere inside neutron stars. The observable leverage is
not uniform across the stellar sequence. GW170817 probes the canonical
mass scale, while the higher-mass \textit{NICER} pulsars determine
whether matter has already crossed into the stiffening region. In that
sense, current observations constrain the \emph{entry} into a
supra-conformal regime more robustly than they constrain the full
rise--peak--fall morphology of the sound-speed profile.

An important caveat is that the inferred peak height is an EOS-model
parameter, not a direct observable. Our smooth baseline family cannot
represent a discontinuous Maxwell-construction first-order transition,
and the transition-capable comparison shows explicitly that allowing a
strong softening dip weakens the supra-conformal inference.
Accordingly, the cleanest robust conclusion from present data is not that
a specific microscopic mechanism is established, but that smooth EOS
families consistently place the onset of dense-matter stiffening in the
mass range of observed heavy neutron stars.

That distinction also clarifies what current data do and do not exclude.
Within smooth EOS families, the posterior support for supra-conformal
behavior around $3$--$4\,\nsat$ is stable across different likelihood
implementations and different smooth parameterizations. Once one permits
strong transition-like softening, however, the evidence no longer
selects decisively between a smooth peaked interpretation and a
transition-capable one. The present data therefore argue more strongly
for the \emph{location} of the relevant density regime than for the
microscopic origin of the stiffening itself.

From that perspective, the most valuable near-term measurements are not
generic additional neutron-star observations, but targeted constraints on
objects that can separate onset from peak access. A modest increase in
the number of well-measured stars above $1.9\,\Msun$ would immediately
test whether the current posterior picture is correct. Likewise,
third-generation gravitational-wave detectors will improve the mapping
between tidal deformability and the density at which the sound speed
begins its rise, but they will be most powerful when combined with new
high-mass radius measurements.
This suggests a practical way to phrase future EOS constraints. A
probability quoted at a chosen density is compact, but a threshold mass
is more directly falsifiable because new observations can immediately
show whether the relevant stars have crossed into the inferred regime.
The present analysis therefore turns a statement about dense-matter
physics into an observational program.

The immediate implication is clear: the most valuable future
measurements are accurate mass--radius constraints for additional stars
between roughly $1.8$ and $2.2\,\Msun$, together with improved tidal
information from nearby binary neutron-star mergers. Those observations
will test whether the currently observed high-mass stars merely enter the
stiffening regime or begin to map its peak directly. In that sense, the
central result of this work is not just a probability that
$\cstwo > 1/3$, but an observational roadmap: current data already point
to the stellar-mass range that must be measured to resolve the next
dense-matter question.

This perspective also helps interpret why present multimessenger
constraints appear stronger in mass space than in density space alone.
The posterior on $\nrise$ is broad when expressed only as a density, but
once translated through exact stellar-structure solutions it maps onto a
much more intuitive observational statement: stars near the canonical
$1.4\,\Msun$ scale rarely reach the inferred onset, whereas stars in the
J0740 class frequently do. The current evidence for stiffening is
therefore not driven by a single abstract density point; it is driven by
the fact that the heaviest observed neutron stars sit precisely in the
mass range where the posterior predicts entry into the stiffening
regime. That mapping is the main reason the result is astrophysically
useful even though its model dependence remains visible in the
cross-family comparison.

The present letter therefore identifies a concrete next step for
multimessenger EOS inference. If future \textit{NICER}-like observations
and next-generation tidal measurements show that several stars between
$1.9$ and $2.2\,\Msun$ consistently exceed the threshold mass
$M_{\rm rise}$ while beginning to approach $M_{\rm peak}$, then the case
for a broad intermediate-density stiffening episode will tighten
substantially. If, instead, improved high-mass measurements remain
compatible with EOS families that delay or interrupt that rise, then the
current smooth-family interpretation will weaken. Either outcome would
be informative. What already emerges from current data is that the
relevant question is no longer simply whether dense matter may become
supra-conformal somewhere, but which observed stars have begun to probe
that regime directly.

This reframing is the main reason the present result is useful beyond the
specific value of any one posterior probability. A statement such as
$P(\cstwo>1/3\ \mathrm{at}\ 3.5\,\nsat)$ is compact, but it is not by
itself an observational target. By contrast, the threshold masses
$M_{\rm rise}$ and $M_{\rm peak}$ identify which pulsars and mergers can
actually test the inference. The existing dataset already reaches the
onset regime through the heaviest observed stars, while only beginning
to touch the peak-access regime.

This is why the present four-family comparison is more informative than
the baseline headline probability alone. The smooth peaked, monotonic,
and piecewise families all place the decisive density interval inside
the mass range of observed heavy neutron stars, even though they assign
different detailed probabilities to supra-conformal behavior. The
transition-capable family weakens the signal precisely because it allows
the stars to encounter a softening interval before any broad stiffening
episode is completed. In practice, that means future data will not have
to distinguish between every microscopic EOS possibility at once. They
will first test a simpler question: do the heaviest measured stars map
onto the onset-access pattern favored by the current posterior?

That question is observationally sharp. If additional stars near
$2\,\Msun$ show radii and tidal responses consistent with matter that has
already crossed into the inferred regime, then the case for a broad
intermediate-density stiffening episode will tighten. If instead the
high-mass population remains consistent with delayed onset or strong
transition-like softening, the present smooth-family interpretation will
weaken. Current multimessenger observations have therefore moved the EOS
problem from a broad statement about dense-matter possibilities to a
direct program of stellar-mass threshold tests.

\begin{acknowledgments}
This work was partially funded by the Millennium Institute of
Subatomic Physics at the High-Energy Frontier, ICN2019\_044.
\end{acknowledgments}

\end{document}